\documentclass[aps,prl,reprint,twocolumn,superscriptaddress,longbibliography,nofootinbib,floatfix,showpacs]{revtex4-1}
\usepackage{epsfig}
\usepackage{amsmath,amssymb,amsfonts}
\usepackage{hyperref}
\usepackage{mathrsfs}
\usepackage{bbm}
\usepackage{slashed}
\usepackage{graphicx}
\usepackage{verbatim}
\usepackage[normalem]{ulem}
\usepackage[usenames]{color}

\newcommand{\be}{\begin{equation}}
\newcommand{\ee}{\end{equation}}
\newcommand{\bi}{\begin{itemize}}
\newcommand{\ei}{\end{itemize}}
\newcommand{\bea}{\begin{eqnarray}}
\newcommand{\eea}{\end{eqnarray}}

\newcommand{\dis}{\displaystyle}
\newcommand{\ud}{\mathrm{d}}
\newcommand{\z}{\boldsymbol{z}}		% vector z
\newcommand{\E}{\mathbf{E}}
\newcommand{\B}{\mathbf{B}}

\def\lambdabar{\protect\@lambdabar}
\def\@lambdabar{%
\relax \bgroup
\def\@tempa{\hbox{\raise.73\ht0
\hbox to0pt{\kern.2\wd0\vrule width.7\wd0
height.1pt depth.1pt\hss}\box0}}%
\mathchoice{\setbox0\hbox{$\displaystyle\lambda$}\@tempa}%
{\setbox0\hbox{$\textstyle\lambda$}\@tempa}%
{\setbox0\hbox{$\scriptstyle\lambda$}\@tempa}%
{\setbox0\hbox{$\scriptscriptstyle\lambda$}\@tempa}%
\egroup }
\newcommand{\red}[1]{{\textcolor{red}{#1}}}  %% RED
\begin{document}

\title{Probing nonperturbative QED with optimally focused laser pulses}

\author{A.~Gonoskov}
\email[]{arkady.gonoskov@physics.umu.se}
\affiliation{Department of Physics, Ume\aa\ University, SE-90187 Ume\aa, Sweden}
\affiliation{Institute of Applied Physics, Russian Academy of Sciences, Nizhny Novgorod 603950, Russia}
\author{I.~Gonoskov}
\email[]{ivan.gonoskov@physics.umu.se}
\affiliation{Department of Physics, Ume\aa\ University, SE-90187 Ume\aa, Sweden}
\author{C.~Harvey}
\email[]{christopher.harvey@qub.ac.uk}
\affiliation{Centre for Plasma Physics, Queen's University Belfast, BT7 1NN, UK}
\author{A.~Ilderton}
\email[]{anton.ilderton@chalmers.se}
\affiliation{Department of Physics, Ume\aa\ University, SE-90187 Ume\aa, Sweden}
\affiliation{Department of Applied Physics, Chalmers University of Technology, SE-41296 Gothenberg, Sweden}

\author{A.~Kim}
\email[]{kim@ufp.appl.sci-nnov.ru}
\affiliation{Institute of Applied Physics, Russian Academy of Sciences, Nizhny Novgorod 603950, Russia}

\author{M.~Marklund}
\email[]{mattias.marklund@physics.umu.se}
\affiliation{Department of Physics, Ume\aa\ University, SE-90187 Ume\aa, Sweden}
\affiliation{Department of Applied Physics, Chalmers University of Technology, SE-41296 Gothenberg, Sweden}
\author{G.~Mourou}
\email[]{gerardmourou@gmail.com}
\affiliation{Institut de la Lumi$\grave{e}$re Extr$\hat{e}$me, ENSTA, Palaiseau, France}
\affiliation{University of Nizhny Novgorod, Nizhny Novgorod 603950, Russia}
\author{A.~Sergeev}
\email[]{ams@ufp.appl.sci-nnov.ru}
\affiliation{Institute of Applied Physics, Russian Academy of Sciences, Nizhny Novgorod 603950, Russia}

\begin{abstract}
We study nonperturbative pair production in intense, focused laser fields called `\textit{e}-dipole' pulses. We address the conditions required, such as the quality of the vacuum, for reaching high intensities without initiating beam-depleting cascades, the number of pairs which can be created, and experimental detection of the created pairs. We find that $e$-dipole pulses offer an optimal method of investigating nonperturbative QED.
\end{abstract}
\maketitle
Understanding quantum field theory in the nonperturbative regime remains a challenging theoretical and experimental issue. Recent advances in technology have spurred interest in the possibility of using intense lasers to probe quantum vacuum phenomena such as nonperturbative electron-positron pair production~\cite{Dunne:2008kc,DiPiazza:2011tq}, a process which is strongly suppressed below the Sauter-Schwinger limit $E_S = m^2c^3/e\hbar\simeq 10^{18}$ V/m \cite{Sauter:1931zz,Schwinger:1951nm}. While field-strengths of this scale are typical of QED, creating them on a macroscopic (laboratory) scale will remain out of our reach for the foreseeable future. Various mechanisms for stimulating pair production have therefore been proposed; these include colliding an intense laser pulse with high energy photons \cite{Baier:2009it}, electrons \cite{Bamber:1999zt}, or other laser pulses~\cite{Schutzhold:2008pz,Dunne:2009gi,Bulanov:2010ei}. The number of pairs which can be created in the collision of laser pulses is sensitive both to the field amplitude \cite{Bulanov:2004de} and the field structure~\cite{Bulanov:2010gb}. In simulating intense laser-matter interactions it is therefore necessary to employ realistic field models.

A potential obstacle to reaching the Sauter-Schwinger limit, or high intensities in general, was raised in~\cite{Fedotov:2010ja}, see also~\cite{Elkina:2010up}. In an experiment, the presence of stray particles due to  imperfect vacuum can result in an avalanche of pair production which is (perturbatively) triggered when particles/photons are dragged/emitted into regions of high strength field. The resulting beam depletion~\cite{NERUSH}, or beam scattering from an emerging electron-positron plasma, then reduces the beam intensity. Even when the effect on the laser radiation is small, generated particles could hinder observation of nonperturbative effects by producing a background which swamps the signals of interest. Hence it is important to understand the conditions leading to cascades, and what backgrounds they produce.

Rather than focusing only on the number of pairs, in this paper we will analyse several aspects of potential pair production experiments.  Our results include a thorough discussion of scenarios where cascades can be avoided, allowing intensities to be raised high enough for nonperturbative pair production to be experimentally observed.

As we explain below, a feasible pulse configuration for next generation laser facilities is an `\textit{e}-dipole' pulse~\cite{Ivan}. These are exact, closed form solutions of Maxwell's equations in vacuum. They exhibit optimal focusing efficiency, (highest possible peak field-strength for a given input power \cite{Bassett}) and describe genuine pulses, having finite energy and finite extent in all four directions. 

This paper is organised as follows. We begin by introducing $e$-dipole pulses and compare their focussing efficiencies with other pulse configurations. We then analyse particle motion in such pulses, estimating the level of vacuum required to keep particles away from, and to limit generation of hard photons into, the focus. We then calculate the number of pairs which could be produced. The behaviour of electron-positron pairs post-creation is then analysed in some detail, before we conclude.

\paragraph{Dipole pulses:--}
There is a limit to the focusing efficiency of a given laser system. Consider first an unfocused, broad laser pulse which can be regarded as nearly monochromatic. The peak field strength $E_0$ which can be obtained by focusing monochromatic light of wavelength $\lambda$ and cycle-averaged power $P$ is bounded above; defining $\mathbb{P}= P/(4\pi\epsilon_0 c)$, the peak field obeys~\cite{Bassett}
\be\label{E0-MAX}
	E_0 \leq \frac{8\pi}{\sqrt{3}}\, \frac{\sqrt{\mathbb{P}}}{\lambda} \simeq 14.51\, \frac{\sqrt{\mathbb{P}}}{\lambda}\;.
\ee
The focused fields which saturate this upper bound are called $e$-dipole pulses, see \cite{Ivan} for full details. $E$-dipole pulses take their name from structural similarities with dipole fields, but do not contain singularities.  They describe a converging pulse of light, with the ideal case of $4\pi$ focusing. The formation of this focused pulse can be pictured as the reverse process of emission from a dipole antenna. The \textit{e}-dipole pulse has the following form. Let $R^2 = x^2+y^2+z^2$ and define the vector $\mathbf Z$ by
\be\label{Z-DEFN}
	\mathbf{Z} = \hat{\z}\,\frac{d}{R}  \big[{g(t+R/c)}-{g(t-R/c)}\big] \;,
\ee
in which the `driving function' $g$ is arbitrary and the `virtual dipole moment' $d$ is a constant. Both $g$ and $d$ can be related to input laser parameters,   see below, because \textit{e}-dipole pulses can be generated by focusing laser fields~\cite{Ivan}.  The electromagnetic fields of an \textit{e}-dipole pulse are then given in terms of $\mathbf Z$ by
\be\label{F-DEFN}
	\E = -\nabla\times\nabla\times {\mathbf{Z}}, \quad \B = -\frac{1}{c^2}\nabla\times \dot{\mathbf{Z}} \;.
\ee
These fields are exact solutions of Maxwell's equations in vacuum. It is easily confirmed that there is no singularity at $R=0$, the focus point. Far from the focus, the electric and magnetic field amplitudes of the \textit{e}-dipole pulse are proportional to $dR^{-1}\ddot{g}\big(t\pm\frac{R}{c}\big)$, and have an angular distribution proportional to $\sin^{2}\theta$, with $\theta$ the polar angle. In the focus one has $\E(0,t)=\hat{\z}\frac{4d}{3c^{3}}\dddot{g}(t)$, $\B(0,t)={0}$. In general we are interested in pulses with, say, Gaussian frequency spreads as for the following driving function
\be\label{QUASI-GAUSS}
	g(\tau) = e^{-(\tau^2/D^2)\ln 4}\sin(\omega \tau) \;,
\ee
in which $\omega$ is the central frequency and $D$ is the full-width-half-maximum duration (i.e.\ the intensity $\sim E^2$ drops to 1/2 its peak value at $t=\pm D/2$). We call such a pulse `quasi-Gaussian' since, far from the focus, the envelope $\ddot g$ has the same Gaussian frequency spread as~$g$. Such pulses are, from (\ref{F-DEFN}), compactly supported in all four directions, which is one advantage of $e$-dipole pulses over other models in the literature; they describe genuine pulses without sacrificing Maxwell's equations. 

The virtual dipole moment $d$ can be expressed as a function of input energy, or power, by using energy conservation in the far-field~\cite{Ivan}. For example, for a monochromatic driving field, frequency $\omega$, the peak focused field strength $E_0$ and $d$ are related by $E_0 = \frac{4}{3}\frac{\omega^3}{c^3} d$. This leads to a simple measure for comparing different pulse models. We define the focusing efficiency parameter $f_E$ via $E_0 = f_E  \frac{\sqrt{\mathbb{P}}}{\lambda}$; this relates the incoming power $P$ to peak field-strength $E_0$. In order to compare with other models, this definition applies to fields which are monochromatic in the far field region. In Table~\ref{FOCUSSING-TABLE} we compare the focusing parameters for three field configurations: two colliding Narozhny-Fofanov beams~\cite{NF,Two}, the theoretical maximum from Fedotov's solution~\cite{Fedotov}, and the $e$-dipole (with monochromatic $g$). The \textit{e}-dipole pulse is most efficient.

Given a pulse model, one can increase (decrease) the focal electric (magnetic) field-strength by colliding several such pulses \cite{Bulanov:2010ei}. This increases the total focusing efficiency and is why, in Table~\ref{FOCUSSING-TABLE}, we compare with two colliding pulses from \cite{NF}, as advocated for pair production in \cite{Two}. Using two pulses eliminates the focal magnetic field, and increases the peak electric field by a factor of $\sqrt{2}$. This `$\sqrt{n}$ increase in focal field strength from $n$ pulses' is limited by 1) the constraint that the total angle of incoming radiation cannot exceed $4\pi$, and  2) imperfect interference in the vector sum of the individual beams' fields, which results from the relative orientations of the incoming beams. The $e$-dipole pulse represents the ideal case of $4\pi$ focussing and maximises the focal electric field strength via optimisation of the polarisation and angular distribution of incoming radiation.

Mimicking an $e$-dipole pulse therefore presents an optimal design for experimental facilities with multi-beam architecture. Our calculations show (details to appear in~\cite{US}) that a realistic configuration of 12 beams with circular apertures, properly arranged and synchronised to imitate the converging $e$-dipole wavefront, provides a field strength just 10\% less than the theoretical maximum of the $e$-dipole pulse. The beam alignment required can be achieved via the reflection of several co-directional beams from a parabolic mirror (similar to the setup in \cite{Ivan}), and synchronisation methods have recently been described in~\cite{mourou.nature.photonics}. Note that the use of multiple laser channels is well-established at projects such as NIF, and will be implemented at next-generation laser facilities such as ELI and XCELS~\cite{ELI,XCELS}.

%%%%%%
\begin{table}[t!]
\begin{tabular}{c || c | c | c}
& Two NF beams \cite{NF,Two} & Fedotov \cite{Fedotov} & \textit{e}-dipole \\  \hline\hline
$ f_E$ &\ $ \dis8\pi\Delta \to 2.51\ $ &\ $\dis 2\pi \sqrt{\frac{5}{3}}\simeq 8.11\ $ &\ $\dis\frac{8\pi}{\sqrt{3}} \simeq 14.51$
\end{tabular}
\caption{\label{FOCUSSING-TABLE} Focussing efficiency $f_E$ (input power to peak electric field) for three beam configurations. We use $\Delta=0.1$ to compare with the literature ($\Delta\ll{1}$ is required \cite{NF,Two,Bulanov:2004de}).}% The $e$-dipole pulse is most efficient.}
\end{table}
%%%%%%%%%%%%%%%%%%%%%%%%

\paragraph*{Interaction with particles within the chamber:--}
As mentioned in the introduction, a potential barrier to the observation of nonperturbative pair production is the background generated by the interactions between the laser and stray particles in the imperfect vacuum of an experimental chamber. If, though, we could guarantee that within the focal volume there were no high energy electrons and photons initiating perturbative processes (or rather, that such events had a low probability), then we could focus beams to the intensities required to trigger nonperturbative pair creation, without first initiating cascades.

To investigate this prospect, one can determine the volume of space from which electrons can initiate cascades, as a function of the initial electron density (initial level of vacuum), and then require that this volume contain less than one particle. We therefore carried out a simulation  of a large number of initially uniformly distributed electrons moving (with radiation reaction accounted for by the Landau-Lifshitz equation~\cite{LL}) in the fields of a converging $e$-dipole pulse, wavelength 810~nm, pulse duration 30~fs (consistent with candidate Ti-sapphire technology for future laser facilities \cite{ELI,XCELS}) and total power $P = 1000$~PW (as is estimated, below, to be required for pair creation, see also Table~\ref{TABLE1}). Laser experiments are now routinely performed in technical vacua of stray molecule density $10^{-8}$--$10^{-9}$mbar, and lower pressures are accessible. We will assume an initial density of $10^5$cm$^{-3}$, equivalent to a pressure of $10^{-12}$mbar. This figure is at the limit of what is achievable today, so is a reasonable assumption for future facilities. The incoming laser pulse will ionise stray molecules and produce electrons. Now, we are interested only in whether electrons are pushed into the focus or not, based on their initial position. Thus, for the determination of the required vacuum quality, we can associate every molecule with a single electron, independent of the level of ionisation. We therefore take the initial electron density to be $10^5 $cm$^{-3}$.

\begin{figure}
\includegraphics[width=0.9\columnwidth]{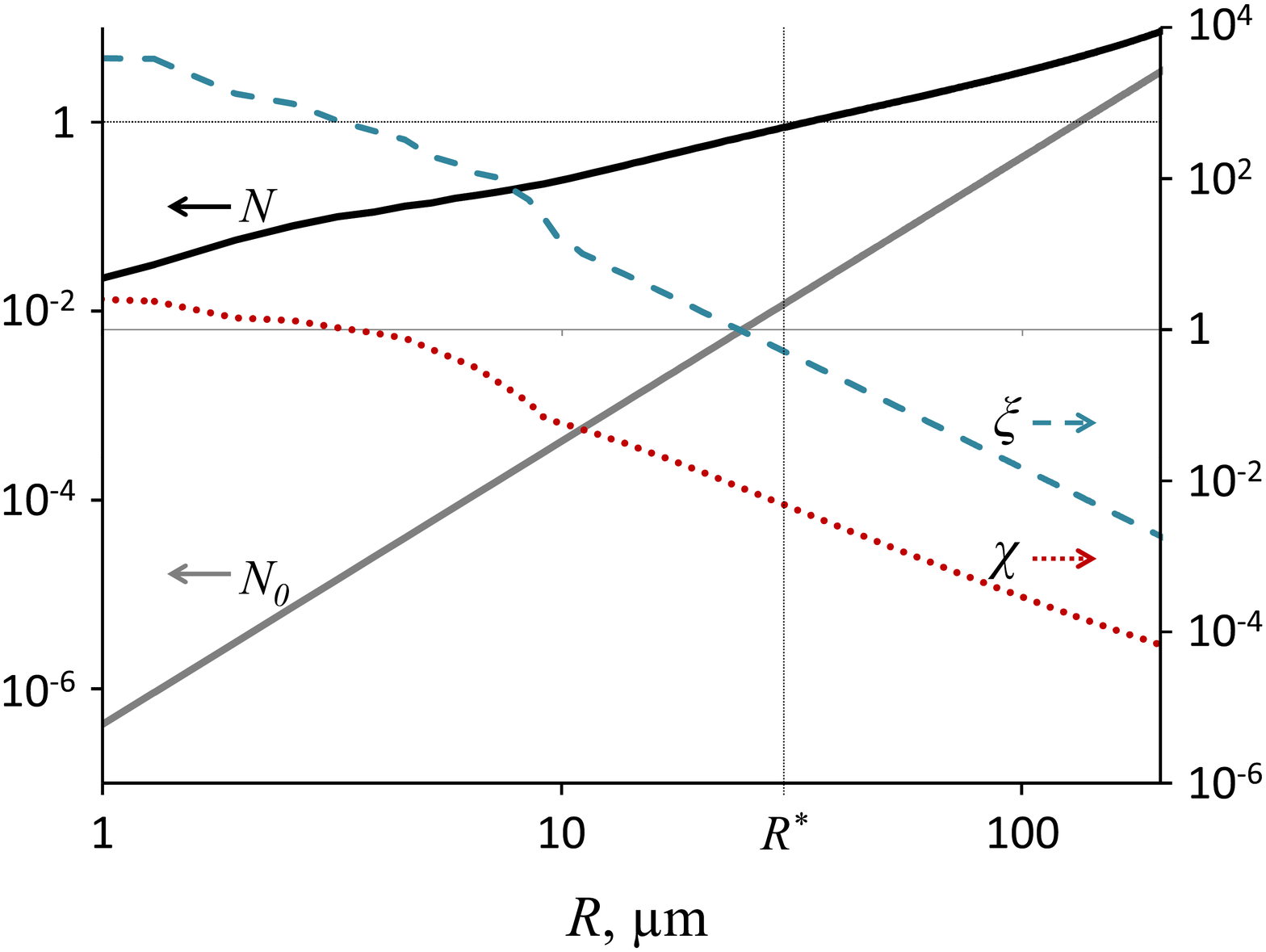}
\caption{\label{compression} Simulation results for electron motion under the influence of an $e$-dipole pulse. {\it Left scale}: initial number ($N_0$) and final average number ($N$) of electrons dragged into a sphere of radius $R$ about the focus, as a function of $R$. {\it Right scale}: maximum values of $\chi$ and $\xi$, taken over all electrons outside the sphere with radius~$R$.} 
\end{figure}

The results of our simulation are shown in Fig.~\ref{compression}. We show here the average number of electrons initially distributed within ($N_0$, calculated from the initial density), or dragged into ($N$), a sphere of radius $R$ around the focus, as a function of $R$.  We see that for the chosen initial density, the final number $N$ of particles drops below one at a radius of around $R^* = 30\mu$m; this is much larger than the focus of the dipole pulse.

We immediately perform two checks on this result. First, we should only base our conclusions on data from regions of space in which quantum effects are negligible. Second, though particles remain {\it outside} the focus, they may still emit hard photons capable of initiating cascades. We therefore tracked in our simulation two parameters, $\chi$ and $\xi$. The first, $\chi$, is the quantum efficiency parameter of the particles~\cite{RitusReview}, calculated along their orbits. This parameter estimates the importance of quantum effects and is, for a particle of momentum $p_\mu$ in a field with energy-momentum tensor $T_{\mu\nu}$~\cite{Heinzl:2008rh,RitusReview}
\be\label{chi-def}
	\chi = \frac{e\hbar}{m^3c^4}\sqrt{p_\mu T^{\mu\nu} p_\nu} \sim \gamma\frac{E}{E_S}\;,
\ee
with $\gamma$ the electron gamma factor. As $\chi$ approaches unity, quantum effects become important. The second parameter, $\xi$, estimates the energy of emitted photons, in ratio to twice the rest mass of an electron. Recall that the typical energy $\hbar \omega_s$ of photons emitted by an ultra-relativistic electron is given in the synchrotron approximation by $\omega_s = \gamma^2(eE /m c)$~\cite{LL}. We therefore define $\xi= \frac{\hbar \omega_s}{2m c^2} \approx \frac{1}{2}\gamma\chi$. For $\xi\ll 1$ we should be able to neglect the emission of hard photons with energy approaching~$2mc^2$.

Returning to our simulation, Fig.~\ref{compression} also shows the {\it maximum} value (over all electrons) of $\chi$ and $\xi$ {\it outside} the sphere of radius $R$. We conclude first that the classical description is valid ($\chi \ll 1$) at least outside the sphere of radius $R^*$, so our estimate of the density increase within this sphere (due to dragging from outside it) can be trusted. Second, since $\xi<1$, particles outside the sphere do not generate photons suitable for initiating pair creation within the sphere; the constant field approximation~\cite{RitusReview}, suitable for very intense fields, shows that the photon-to-pair decay time is, at $\xi<1$, orders of magnitude longer than the pulse duration. Hence we conclude from Fig.~\ref{compression} that an initial density of less than $10^5$~cm$^{-3}$ leads to an average of well less than one particle existing in or entering into the sphere of radius $R^*$, and an absence of photons sufficiently hard to create pairs within that sphere. Our estimate for the required initial density is overly cautious: more accurate estimates which include data on, e.g.\ the emission direction of radiation could significantly reduce the stated requirements. (Our simulations show that hard photons are typically emitted away from, rather than toward, the focus.)

\paragraph{Pair production:--}
%%%%%%%%%
%
The number of pairs which can be created in a given field depends on the local Lorentz invariants~\cite{Schwinger:1951nm}
\be
	\mathcal{S}=-\frac{1}{4}F_{\mu\nu}F^{\mu\nu}\quad \text{and}\quad	\mathcal{P}=-\frac{1}{4}{\tilde F}_{\mu\nu}F^{\mu\nu} \;.
\ee
In an \textit{e}-dipole pulse we have $\mathcal{P}=\E\cdot\B ={0}$, i.e.\ the electric and magnetic fields are orthogonal in all space. The invariant $\mathcal S=(\E^2-\B^2)/2$ can be positive (the electric field `dominates') and the pulse is capable of pair production~\cite{Sauter:1931zz,H-E,Schwinger:1951nm}. To maximise the pair production rate, one must maximise the relevant invariant, i.e.\ maximise the electric field and eliminate the magnetic field in the focus. This is achieved (as above) by the use of counter-propagating pulses \cite{Two,Bulanov:2010ei}. Since the $e$-dipole pulse has $\B=0$ in the focus and yields the optimal focal electric field, it also gives the most efficient `conversion' of input power into invariant $\epsilon$, i.e.\ allows efficient conversion of energy into produced pairs.

To calculate the number of produced pairs, we use the locally-constant-field estimate~\cite{Bulanov:2004de}, based on \cite{H-E}. For a field in which $\mathcal{P}\equiv 0$ we have 
\be\label{N}
	N_\text{pairs} = \frac{1}{4\pi^3 \lambdabar_c^4}\int\!\ud^4x \ \epsilon^2(x) \exp\bigg[-\frac{\pi}{\epsilon(x)}\bigg] \;,
\ee
where $\epsilon= \sqrt{\mathcal{S}+|\mathcal{S}|}/E_S$, and $\lambdabar_c = \hbar/mc$ is the reduced Compton wavelength. Since the main contribution to $N_\text{pairs}$ comes from the focal region (which in a quasi-Gaussian is localised both in space and time), and in order to compare with other pulse models (which are typically monochromatic in the far-field), we temporarily drop the envelope in (\ref{QUASI-GAUSS}), setting $D=\infty$, and quote only the contribution to $N_\text{pairs}$ from a single cycle.

%%%%%%%%%%%%%%%%%%%%%%
\begin{table}[t!]
\begin{tabular}{l||c|c|c}
& $\lambda =1\, \mu$m\    & $\lambda =0.8\, \mu$m\ & $\lambda = 0.4\, \mu$m \\ \hline\hline
$1660$ PW\ &\ 1 \red{\scriptsize ($5.5$ kJ)} &\ $10^3$ \red{\scriptsize ($4.4$ kJ)} &\ $10^{10}$ \red{\scriptsize ($2$ kJ)} \\ \hline
$1120$ PW\ &\ $10^{-4}$ \red{\scriptsize ($3.7$ kJ)} &\ $1$ \red{\scriptsize ($3$ kJ)}&\ $10^8$ \red{\scriptsize ($1.5$ kJ)}  \\ \hline
$320$ PW\ &\ $10^{-23}$ \red{\scriptsize ($1.1$ kJ)} &\ $10^{-14} $\red{\scriptsize ($0.85$ kJ)} &\ $1$ \red{\scriptsize ($0.43$ kJ)}  
\end{tabular}
\caption{\label{TABLE1} The number of pairs produced (order of magnitude) as a function of wavelength and power. Bracketed figures give the incoming energy in one cycle. Parameters are chosen so that the diagonal values exhibit the threshold (average) power required to create a single pair in one cycle.}
\end{table}
%%%%%%%%%%%%%%%%%%%%%%
%
While the number of created pairs depends on the invariant $\epsilon$, the relevant experimental quantity is the total available input energy. The energy required to reach a given peak $\epsilon$ will be minimal when using an $e$-dipole pulse configuration, because of the pulse's optimal focusing efficiency. The number of pairs produced is shown in Table~\ref{TABLE1}, as a function of the driving field's wavelength, (average) input power and energy. Our results demonstrate that the threshold power required for pair creation is around $P=1120$\,PW (at a wavelength of $\lambda=0.8\mu$m), requiring only a few kJ of energy. The corresponding peak electric field at threshold is $E_0\simeq 0.08 E_S$, agreeing with previous estimates~\cite{Fedotov,Bulanov:2010ei}.

All predictions of the pair yield based on (\ref{N}) should of course be interpreted as a measure of how easy it is to {\it produce} a pair, given an initial power and frequency, rather than a number of pairs. The reason for this is that once a pair is created, other processes can occur, either increasing or decreasing the net number of pairs which can be observed. In the former category is cascade formation \cite{Bell:2008zzb, Elkina:2010up}, in which the pairs are accelerated and emit hard photons which create further pairs, and so on~\cite{Fedotov:2010ja,Bulanov:2010gb}. In the latter category are processes such as pair annihilation, though the annihilation cross section in intense fields  is typically much smaller than that of cascade-generating processes~\cite{RitusReview,Voroshilo:2010zz}. It is therefore important to address how one might best detect pair production events. We turn to this now, by analysing the behaviour of the created pairs.

\begin{figure}[t!]
		\includegraphics[width=0.5\columnwidth]{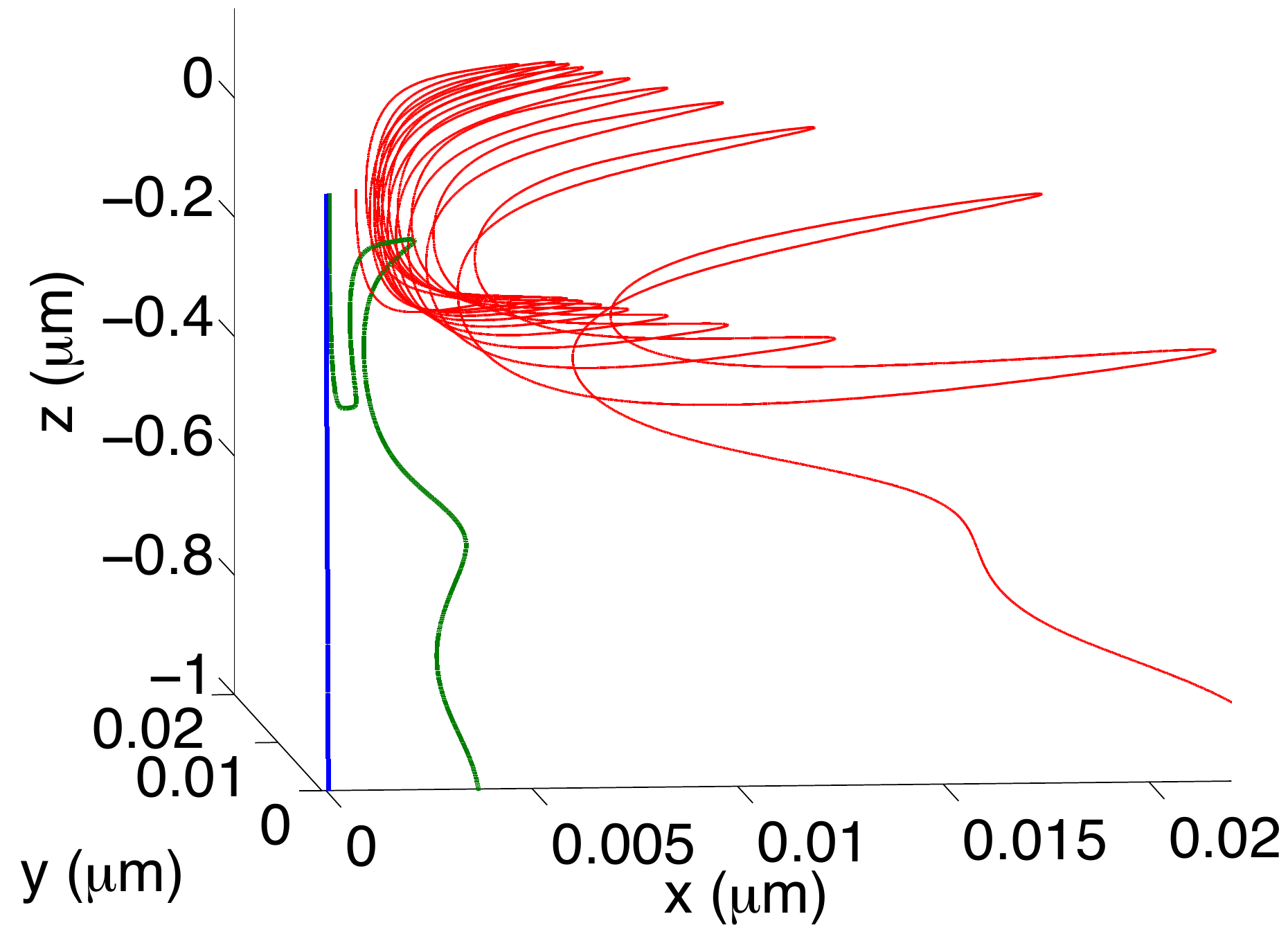}\includegraphics[width=0.5\columnwidth]{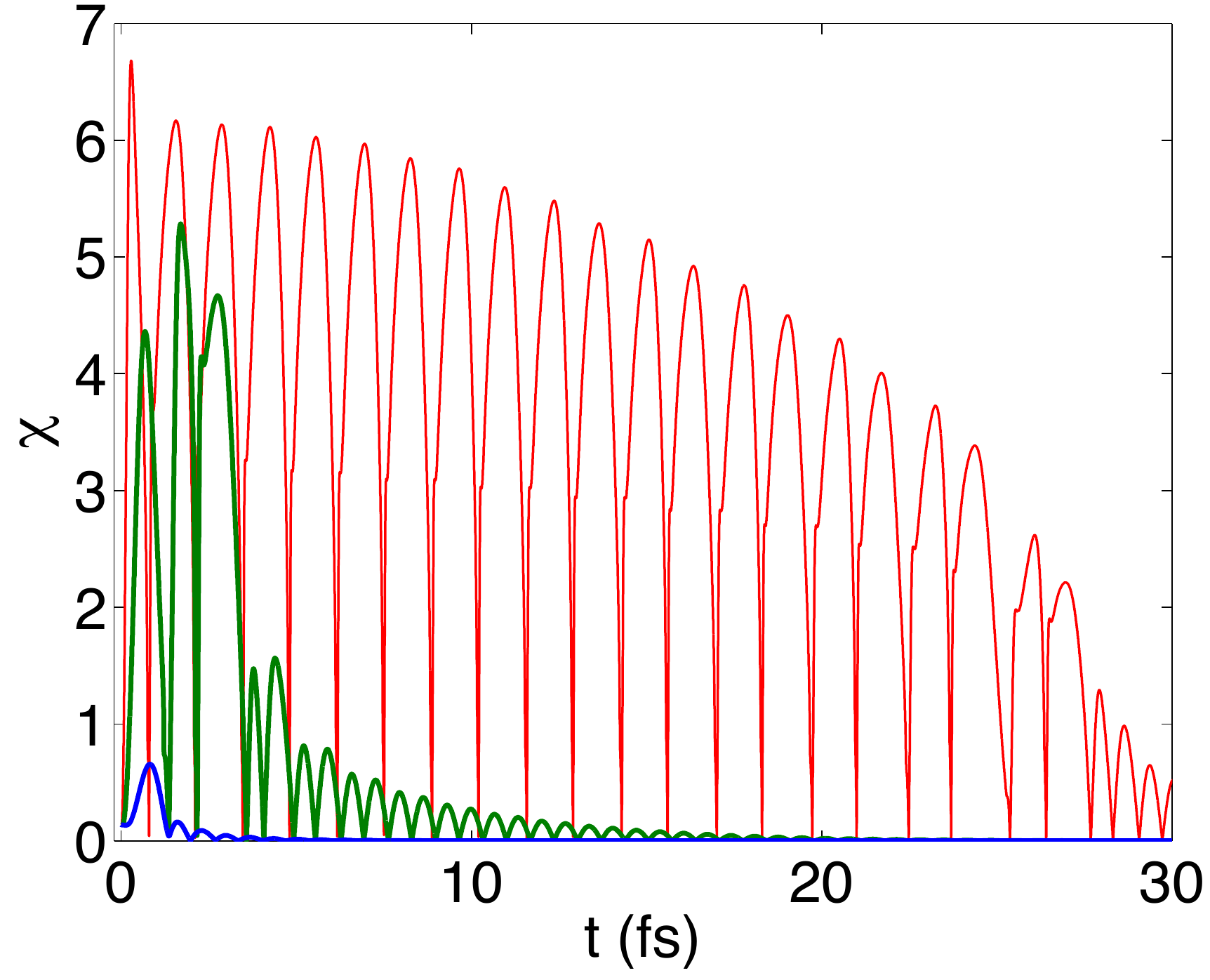}
		\caption{\label{30fs_plots} Trajectories [left] and $\chi$ parameter [right] for particles created in a 30~fs, 1100 PW pulse. Initial positions: $x=y=z=10^{-3}\lambda$ (red), $x=y=z=10^{-4}\lambda$ (green), $x=y=z=10^{-5}\lambda$ (blue).}
\end{figure}

\paragraph{Post-creation behaviour:--} Fig.~\ref{30fs_plots} shows typical trajectories, and $\chi$ values, for positrons born at rest near the focus of a $D=30$ fs $e$-dipole pulse. The trajectories depend sensitively on where the particles are created. The majority of particles, with a typical trajectory shown in red in Fig.~\ref{30fs_plots}, have large $\chi$-values, implying that the depicted motion will receive significant quantum corrections. However, once the particles leave the focus, their $\chi$ drops quickly to below one and classical predictions become accurate.  Our simulations show that radiation reaction is responsible for recirculating particles back into the focus, and has an added effect of keeping the particles `in phase' with the field intensity gradient, which is the cause of their large $\chi$ values. Recirculation returns the particles to the quantum regime, increasing the possibility of additional QED processes occurring, as predicted in \cite{Elkina:2010up}.

Close to the focus, the longitudinal electric field dominates over the other field components, causing particles born in this region to simply exit the pulse parallel to the $z$-axis, without oscillation. (See~\cite{Bulanov:2010gb} for related behaviour.) Our simulations show that these particles are `out of phase' with the field, so that while they have high $\gamma$ factors, they also have $\chi<1$, and so can be analysed classically. These particles will emit radiation in a small cone about the $z$-axis. We therefore suggest the detection of particles exiting in the $z$-direction as a possible candidate for direct measurement of the produced pairs.
\paragraph*{Conclusions:--} We have considered a potential experimental setup for measuring nonperturbative pair production in $e$-dipole pulses. Our analysis of stray particles in the chamber, and their emission, shows that cascade initiation (and resulting beam depletion) can be avoided for sufficient levels of vacuum. This allows higher focal intensities to be reached, and allows us to consider the possibility of experimentally measuring nonperturbative pair creation, for which we estimate a total power of 1000 PW is required.
\acknowledgements
We thank the Swedish National Infrastructure for Computing (SNIC) for supercomputer sources. A.I. thanks C.~Murphy for discussions. The authors are supported by the Ministry of Education and Science of the Russian Federation, Agreement No.\ 11.G34.31.0011 (A.G., G.M., A.S.), the Russian Foundation for Basic Research grant No. 12-02-12086 (A.G.), the Swedish Research Council, contract 2011-4221 (A.I.), the European Research Council contract 204059-QPQV (A.I., M.M.) and EPSRC grant EP/I029206/1--YOTTA (C.H.).


\begin{thebibliography}{99}

%\cite{Dunne:2008kc}
\bibitem{Dunne:2008kc}
  G.~V.~Dunne,
  %``New Strong-Field QED Effects at ELI: Nonperturbative Vacuum Pair Production,''
  Eur.\ Phys.\ J.\ D {\bf 55} (2009) 327
  %[arXiv:0812.3163 [hep-th]].
  %%CITATION = ARXIV:0812.3163;%%

%\cite{DiPiazza:2011tq}
\bibitem{DiPiazza:2011tq}
  A.~Di Piazza {\it et al.},
  %``Extremely high-intensity laser interactions with fundamental quantum systems,''
  Rev.\ Mod.\ Phys.\  {\bf 84} (2012) 1177.
  %[arXiv:1111.3886 [hep-ph]].
  %%CITATION = ARXIV:1111.3886;%%

%\cite{Sauter:1931zz}
\bibitem{Sauter:1931zz}
  F.~Sauter,
  %``Uber das Verhalten eines Elektrons im homogenen elektrischen Feld nach der
  %relativistischen Theorie Diracs,''
  Z.\ Phys.\  {\bf 69}, 742 (1931).
  %%CITATION = ZEPYA,69,742;%%

%\cite{Schwinger:1951nm}
\bibitem{Schwinger:1951nm}
  J.~S.~Schwinger,
  %``On gauge invariance and vacuum polarization,''
  Phys.\ Rev.\  {\bf 82}, 664 (1951).
  %%CITATION = PHRVA,82,664;%%

%\cite{Baier:2009it}
\bibitem{Baier:2009it}
  V.~N.~Baier and V.~M.~Katkov,
  %``Pair creation by a photon in an electric field,''
  Phys.\ Lett.\ A {\bf 374} (2010) 2201.
%  [arXiv:0912.5250 [hep-ph]].
  %%CITATION = ARXIV:0912.5250;%%


%\cite{Bamber:1999zt}
\bibitem{Bamber:1999zt}
  C.~Bamber {\it et al.},
  %``Studies of nonlinear QED in collisions of 46.6-GeV electrons with  intense
  %laser pulses,''
  Phys.\ Rev.\  D {\bf 60}, 092004 (1999).
  %%CITATION = PHRVA,D60,092004;%%
  
  %\cite{Schutzhold:2008pz}
\bibitem{Schutzhold:2008pz}
  R.~Sch\"utzhold, H.~Gies and G.~Dunne,
  %``Dynamically assisted Schwinger mechanism,''
  Phys.\ Rev.\ Lett.\  {\bf 101}, 130404 (2008)
  %[arXiv:0807.0754 [hep-th]].
  %%CITATION = PRLTA,101,130404;%%

  %\cite{Dunne:2009gi}
\bibitem{Dunne:2009gi}
  G.~V.~Dunne, H.~Gies and R.~Schutzhold,
  %``Catalysis of Schwinger Vacuum Pair Production,''
  Phys.\ Rev.\  D {\bf 80}, 111301 (2009)
  %[arXiv:0908.0948 [hep-ph]].
  %%CITATION = PHRVA,D80,111301;%%

  %\cite{Bulanov:2010ei}
\bibitem{Bulanov:2010ei}
  S.~S.~Bulanov, V.~D.~Mur, N.~B.~Narozhny, J.~Nees and V.~S.~Popov,
  %``Multiple colliding electromagnetic pulses: a way to lower the threshold of $e^+e^-$ pair production from vacuum,''
  Phys.\ Rev.\ Lett.\  {\bf 104} (2010) 220404.
%  [arXiv:1003.2623 [hep-ph]].
  %%CITATION = ARXIV:1003.2623;%%

%\cite{Bulanov:2004de}
\bibitem{Bulanov:2004de}
  S.~S.~Bulanov, N.~B.~Narozhny, V.~D.~Mur and V.~S.~Popov,
  %``On e+ e- pair production by a focused laser pulse in vacuum,''
  Phys.\ Lett.\ A {\bf 330} (2004) 1.%; JETP Lett.\ {\bf 80} (2004) 382; JETP {\bf 129} (2006) 14.
  %%CITATION = HEP-PH/0403163;%%
  
  %\cite{Bulanov:2010gb}
\bibitem{Bulanov:2010gb}
  S.~S.~Bulanov, T.~Z.~Esirkepov, A.~G.~R.~Thomas, J.~K.~Koga and S.~V.~Bulanov,
  %``On the Schwinger limit attainability with extreme power lasers,''
  Phys.\ Rev.\ Lett.\  {\bf 105} (2010) 220407.
  %[arXiv:1007.4306 [physics.plasm-ph]].
  %%CITATION = ARXIV:1007.4306;%%

%\cite{Fedotov:2010ja}
\bibitem{Fedotov:2010ja}
  A.~M.~Fedotov, N.~B.~Narozhny, G.~Mourou and G.~Korn,
  %``Limitations on the attainable intensity of high power lasers,''
  Phys.\ Rev.\ Lett.\  {\bf 105} (2010) 080402.
 % [arXiv:1004.5398 [hep-ph]].
  %%CITATION = ARXIV:1004.5398;%%

%\cite{Elkina:2010up}
\bibitem{Elkina:2010up}
  N.~V.~Elkina {\it et al.}, %A.~M.~Fedotov, I.~Y.~Kostyukov, M.~V.~Legkov, N.~B.~Narozhny, E.~N.~Nerush and H.~Ruhl,
  %``QED cascades induced by circularly polarized laser fields,''
  Phys.\ Rev.\ ST Accl.\ Beams {\bf 14} (2011) 054401.
 % [arXiv:1010.4528 [hep-ph]].
  %%CITATION = ARXIV:1010.4528;%%

%\cite{Nerush:2010fe}
\bibitem{NERUSH}
  E.~N.~Nerush, I.~Y.~.Kostyukov, A.~M.~Fedotov, N.~B.~Narozhny, N.~V.~Elkina and H.~Ruhl,
  %``Laser field absorption in self-generated electron-positron pair plasma,''
  Phys.\ Rev.\ Lett.\  {\bf 106} (2011) 035001
   [Erratum-ibid.\  {\bf 106} (2011) 109902]
%  [arXiv:1011.0958 [physics.plasm-ph]].
  %%CITATION = ARXIV:1011.0958;%%
  %18 citations counted in INSPIRE as of 07 May 2013
\bibitem{Ivan}
I.~Gonoskov, A.~Aiello, S.~Heugel, G.~Leuchs, 
%Dipole pulse theory: Maximizing the field amplitude from 4? focused laser pulses
Phys.\ Rev.\ A {\bf 86} (2012) 053836.

\bibitem{Bassett} 
  I.~M.~Bassett,
  Optica Acta: Int.\ J.\ Optics {\bf 33} (1986) 279.

\bibitem{NF}
N.~B.~Narozhny, M.~S.~Fofanov, JETP {\bf 117} (2000) 867.

\bibitem{Two}
S.~S.~Bulanov, N.~B.~Narozhny, V.~D.~Mur and V.~S.~Popov,
JETP {\bf 102} (2006) 9.

\bibitem{Fedotov}
A.\ M.\ Fedotov, Laser Physics 19 (2009) 214.

\bibitem{US}
To appear.

\bibitem{LL}
L.~D.~Landau, E.~M.~Lifshitz, {\it The classical theory
of fields}, 1975, Elsevier, Oxford.

\bibitem{mourou.nature.photonics}
G.~Mourou, B.~Brocklesby, T.~Tajima, J.~Limpert,
Nature Photonics {\bf 7} (2013) 258.
    
\bibitem{ELI} \href{http://www.extreme-light-infrastructure.eu/}{\texttt{http://www.extreme-light-infrastructure.eu/}}.
\bibitem{XCELS} \href{http://www.xcels.iapras.ru}{\texttt{http://www.xcels.iapras.ru}}.

\bibitem{RitusReview}
%
V.~I.~Ritus, %{\it Quantum effects of the interaction of elementary particles with an intense electromagnetic field},
J.~Russian Laser Research {\bf 6} (1985) 497.

%\cite{Heinzl:2008rh}
\bibitem{Heinzl:2008rh}
%The common form with $pF^2p$ instead of $pTp$ applies only to fields with vanishing Lorentz invariants.
T.~Heinzl and A.~Ilderton, Opt.\ Commun.\  {\bf 282} (2009)~1879.
 % [arXiv:0807.1841 [physics.class-ph]].
  %%CITATION = ARXIV:0807.1841;%%
  
    
\bibitem{H-E} 
  W.~Heisenberg, H.~Euler,
  Z. Phys., {\bf 33}, 714, (1936).
  
  %\cite{Bell:2008zzb}
\bibitem{Bell:2008zzb}
  A.~R.~Bell and J.~G.~Kirk,
  %``Possibility of Prolific Pair Production with High-Power Lasers,''
  Phys.\ Rev.\ Lett.\  {\bf 101} (2008) 200403.
  %%CITATION = PRLTA,101,200403;%%

  %\cite{Voroshilo:2010zz}
\bibitem{Voroshilo:2010zz}
  A.~I.~Voroshilo, E.~A.~Padusenko and S.~P.~Roshchupkin,
  %``One-photon annihilation of an electron-positron pair in the field of pulsed circularly polarized light wave,''
  Laser Phys.\  {\bf 20} (2010) 1679.
  %%CITATION = LAPHE,20,1679;%%

\end{thebibliography}
\end{document}